\documentclass{article}
\usepackage{spconf,amsmath,graphicx,hyperref}
\usepackage{amssymb}

\usepackage{subcaption}
\usepackage{multirow}      
\usepackage{makecell}
\usepackage{enumitem}
\usepackage[skip=3pt]{caption}

\hypersetup{
    colorlinks=true,
    citecolor=black,
    linkcolor=black,
    urlcolor=black
}

\title{Shared Representation Learning for Reference-Guided Targeted Sound Detection}
\name{Shubham Gupta\textsuperscript{*}, Adarsh Arigala\textsuperscript{*}, B. R. Dilleswari,   Sri Rama Murty Kodukula
\thanks{\textsuperscript{*}These authors contributed equally. Source code and pretrained models are available at \url{https://github.com/ArigalaAdarsh/Reference-Guided-Targeted-Sound-Detection}.}}
\address{
Speech Information and Processing Lab, \\  Indian Institute of Technology Hyderabad, India \\
guptashubham0318@gmail.com, arigalaadarsh780@gmail.com,  r180909@rguktrkv.ac.in,  ksrm@ee.iith.ac.in
}

\begin{document}
%
\maketitle
\begin{abstract}
Human listeners exhibit the remarkable ability to segregate a desired sound from complex acoustic scenes through selective auditory attention, motivating the study of Targeted Sound Detection (TSD). The task requires detecting and localizing a target sound in a mixture when a reference audio of that sound is provided. Prior approaches, rely on generating a sound-discriminative conditional embedding vector for the reference and pairing it with a mixture encoder, jointly optimized with a multi-task learning approach. In this work, we propose a unified encoder architecture that processes both the reference and mixture audio within a shared representation space, promoting stronger alignment while reducing architectural complexity. This design choice not only simplifies the overall framework but also enhances generalization to unseen classes. Following the multi-task training paradigm, our method achieves substantial improvements over prior approaches, surpassing existing methods and establishing a new state-of-the-art benchmark for targeted sound detection, with a segment-level F1 score of 83.15\% and an overall accuracy of 95.17\% on the URBAN-SED dataset.

\end{abstract}
\section{Introduction}
\label{sec:intro}
Real-world acoustic environments often contain overlapping sound events, making it challenging to detect and localize a specific event of interest. Traditional Sound Event Detection (SED) systems address this by classifying and localizing all events from a fixed set of categories \cite{19,20,21}. However, in applications such as surveillance \cite{15}, multimedia retrieval \cite{16}, and smart assistants \cite{17}, the goal is not exhaustive labeling but retrieval of a user-specified target event. This motivates the task of Targeted Sound Detection (TSD), where the system is conditioned on a reference audio clip and must determine whether the corresponding event occurs  within a longer and potentially noisy mixture. Unlike conventional SED, TSD does not rely on pre-defined class labels, making it more extensible to unseen-class generalization while reducing dependence on large annotated datasets. The task reflects the human ability to selectively focus auditory attention, the ``cocktail party`` effect, where one can focus on a particular sound amidst competing noise.

TSD is closely related to neighboring tasks but remains distinct in its objectives. Target speaker extraction \cite{11,12} focuses on isolating a speaker’s voice from a multi-speaker mixture, given a short reference utterance of the target speaker. Similarly, Target Sound Extraction (TSE) \cite{14} extends this idea to general acoustic events, aiming to reconstruct the waveform of a specified sound class from a mixture when a reference example is provided. In contrast, TSD does not attempt to recover the full signal; it is concerned only with detecting and localizing the occurrence of the target event, a formulation that is more efficient and better aligned with large-scale retrieval scenarios. We mention TSE only to position TSD within the broader family of reference-guided audio tasks; direct comparison with TSE models is beyond the scope of this work and left for future exploration. The main contributions of this work are:

\begin{enumerate}[topsep=2pt,itemsep=1pt,parsep=0pt]
    \item We propose a unified encoder framework for targeted sound detection (TSD) that jointly processes reference and mixture audio in a shared representation space, reducing architectural complexity while improving feature alignment and generalization to unseen classes.
    \item We introduce a flexible reference-mixture fusion framework and systematically evaluate multiple conditioning strategies, including element-wise multiplication, FiLM-based conditioning \cite{perez2017filmvisualreasoninggeneral}, and cross-attention.
    \item We achieve state-of-the-art segment-level detection performance on URBAN-SED and demonstrate robust cross-domain generalization on AudioSet-Strong.
\end{enumerate}

\begin{figure*}[!t]  
    \centering
    \includegraphics[width=\linewidth,height=5cm]{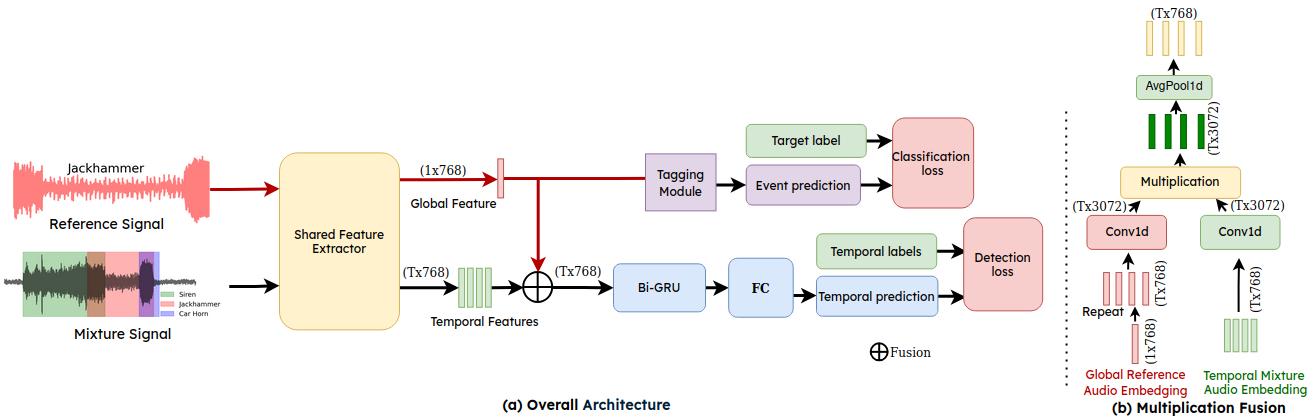}
    \captionsetup{font=small}
    \caption{Overview of the proposed method. The tagging module comprises two fully connected layers followed by a softmax layer for event prediction.}
    \label{fig:architecture} 
\end{figure*}

\section{RELATED WORK}
\label{sec:RELATED WORK}
Early progress in Targeted Sound Detection (TSD) has been driven by neural architectures conditioned on a reference example. One of the first such methods, TSDNet \cite{3}, employs a two-branch design where a conditional network encodes the reference audio into a discriminative embedding, and a detection network fuses this representation with mixture features to predict target activations. Trained on the URBAN-TSD dataset constructed from URBAN-SED and UrbanSound8K, TSDNet demonstrated the feasibility of both strong and weak supervision, achieving segment-level F1 scores of 76.3\% and 56.8\%, respectively. The framework was further enhanced with spectrogram mixup augmentation and multi-task optimization, establishing a baseline for subsequent research.

Building on this foundation, Yang et al. \cite{18} introduced a mixed-supervised approach in which a strongly-supervised and a weakly-supervised model iteratively refine one another. More recently, RaDur \cite{10} addressed practical challenges such as noisy or short reference signals and the detection of transient events and incorporated a reference-aware embedding enhancement module, attention pooling, and a duration-robust focal loss. Despite these advances, the literature on TSD remains sparse, and further progress is needed to develop architectures that generalize effectively to unseen classes and challenging real-world conditions.

\section{Methodology}
\label{sec:Methodology}

We adopt ConvNeXt \cite{9} as a unified audio encoder to extract representations from both the mixture and reference audio. The system comprises three components: (i) a shared ConvNeXt-based audio encoder, (ii) a fusion and temporal modeling module, and (iii) a multi-task loss combining clip-level classification and frame-level detection. An overview of the architecture is shown
in Fig.\ref{fig:architecture}.

\subsection{Audio Encoder} 
We use a single ConvNeXt encoder pre-trained on AudioSet-2M \cite{8} to extract representations from both mixture and reference log–Mel spectrograms. It produces frame-level embeddings $\mathbf{H}_m \in \mathbb{R}^{T \times F}$ for the mixture and a global clip embedding $\mathbf{h}_{\text{ref}} \in \mathbb{R}^{1 \times F}$ for the reference, where $T$ denotes the number of time frames and $F$ the feature dimension. The unified encoder maps both audio representations into a shared embedding space, reducing architectural complexity and promoting stronger reference–mixture alignment.


\subsection{Fusion and Temporal Modeling}
As shown in Fig. \ref{fig:architecture}, the reference embedding $h_{\text{ref}}$ is repeated along the temporal axis to match the mixture frames, yielding $\mathbf{H}_{\text{ref}} \in \mathbb{R}^{T \times F}$. Both $\mathbf{H}_{\text{ref}}$ and $\mathbf{H}_m$ are projected to a common dimension $F'$ using separate 1-D convolutions, $\tilde{\mathbf{H}}_{\text{ref}} = \mathrm{Conv1D}_{\text{ref}}(\mathbf{H}_{\text{ref}})$ and $\tilde{\mathbf{H}}_m = \mathrm{Conv1D}_{\text{mix}}(\mathbf{H}_m) \in \mathbb{R}^{T \times F'}$. The streams are fused via element-wise multiplication and temporal average pooling, $\mathbf{Z} = \mathrm{AvgPool1D}(\tilde{\mathbf{H}}_{\text{ref}} \odot \tilde{\mathbf{H}}_m)$, and passed to a BiGRU for temporal modeling, $\mathbf{H}_{\text{gru}} = \mathrm{BiGRU}(\mathbf{Z}) \in \mathbb{R}^{T \times F}$. A fully connected layer outputs frame-level logits $o_i$ and probabilities $\hat{p}_i = \sigma(o_i)$ for each frame $i$. More expressive fusion schemes, including FiLM \cite{perez2017filmvisualreasoninggeneral} and cross-attention, are evaluated in Section~5.

\subsection{Loss Function}
We jointly optimize the model using clip-level and frame-level objectives. For clip-level classification, the reference embedding $\mathbf{h}_{\text{ref}}\!\in\!\mathbb{R}^{1\times F}$ is passed through a linear layer to obtain logits $\mathbf{z}\!\in\!\mathbb{R}^{C}$ over the target classes ($C\!=\!10$ for URBAN-SED). Given one-hot labels $\mathbf{y}$, the cross-entropy loss is 
$\mathcal{L}_{\text{CE}}=-\sum_{c=1}^{C}y_{c}\log \frac{e^{z_{c}}}{\sum_{c'}e^{z_{c'}}}$.
For frame-level detection, given ground-truth labels $p_{i}\!\in\!\{0,1\}$ and predicted probabilities $\hat{p}_{i}$ at each frame, we minimize the binary cross-entropy loss 
$\mathcal{L}_{\text{SED}}=-\sum_{i=1}^{T}[\,p_{i}\log\hat{p}_{i}+(1-p_{i})\log(1-\hat{p}_{i})\,]$.
The overall objective combines both terms:
\[
\mathcal{L}_{\text{total}}=\mathcal{L}_{\text{CE}}+\mathcal{L}_{\text{SED}}.
\]
This design encourages the reference embedding to supervise coarse class prediction while the frame-level objective drives fine-grained temporal localization of the target sound.
 
\section{Experimental Setup}
\label{sec:experiments}
\subsection{Dataset}
\label{label:dataset}
Experiments are conducted on URBAN-SED \cite{2} and UrbanSound8K \cite{1}, following the TSD construction protocol of Wang et al.~\cite{3}. URBAN-SED contains 10,000 synthetic 10-s urban soundscapes with 1–9 events from 10 classes, split into 6,000/2,000/2,000 train/val/test clips, each with strong onset/offset annotations. UrbanSound8K comprises 8,732 isolated clips of the same classes and serves as the reference audio source.

We derive two benchmarks: \textbf{Urban-TSD-Strong} and \textbf{Urban-TSD-Strong+}. In Urban-TSD-Strong, each sample pairs a URBAN-SED mixture with a reference clip from UrbanSound8K belonging to a class present in the mixture, along with strong target labels. For mixtures containing \(N\) target events, up to \(N\) positive reference–mixture pairs are formed. Urban-TSD-Strong+ augments this setup with negative samples, where the reference class is absent from the mixture, better reflecting realistic query scenarios. Dataset statistics are summarized in Table~\ref{tab:sample_count}.

Cross-domain generalization is evaluated on a curated subset of AudioSet-Strong, comprising real-world YouTube recordings with strong annotations. We retain clips containing at least one of the ten URBAN-SED classes while continuing to use UrbanSound8K references, yielding a more challenging test under distribution shift.
\subsection{Training Configuration}
\label{ssec:Implementation}

All audio signals are resampled to 32\,kHz prior to feature extraction. Log–Mel spectrograms are computed using 224 Mel bins with a window size of 1024 and a hop size of 320. The encoder feature dimension is set to $F=768$, and the projected fusion dimension is set to $F'=3072$. The model is implemented in PyTorch and trained for 30 epochs with a batch size of 16 using the AdamW optimizer with an initial learning rate of $1{\times}10^{-4}$. A Reduce-on-Plateau scheduler decreases the learning rate by a factor of 0.1 after three consecutive epochs without improvement on the validation set. To enhance robustness and mitigate overfitting, spectrogram-level data augmentations including frequency masking, time masking, and temporal shifting are applied during training. The ConvNeXt encoder, pre-trained on AudioSet, is further fine-tuned to adapt its representations to the targeted sound detection task.

 \begin{table}[!t]
\centering
\renewcommand{\arraystretch}{1.0} 
\setlength{\tabcolsep}{4pt}       
\captionsetup{font=small}
\caption{Count of Strong and Strong+ labelled data across splits.}
\small
\begin{tabular}{lcc}
\hline
Type        & Strong & Strong+ \\
\hline
Training    & 23106  & 29106   \\
Validation  & 7681   & 9681    \\
Test        & 7702   & 9702    \\
\hline
\end{tabular}
 
\label{tab:sample_count}
\end{table}

\section{Results}
\label{sec:section5}
We evaluate detection performance using the segment-based F-measure \cite{22}, a standard metric in sound event detection, together with class-wise accuracy as implemented in the \texttt{sed\_eval}\footnote{\url{https://tut-arg.github.io/sed_eval/}} toolbox, using a fixed segment size of 200\,ms. All F-scores and accuracies are macro-averaged across classes. Frame-level predictions are obtained by binarizing posterior probabilities at a threshold of 0.37 (selected on the validation set) and applying a median filter of width three frames to suppress spurious activations. The resulting binary sequence is mapped to onset/offset timestamps based on the encoder frame hop. We clarify that the system operates in a class-conditional TSD setting: multi-class supervision is used only during training, while inference performs a binary presence/absence decision conditioned on the reference.

We benchmark our framework against representative methods including TSDNet \cite{3}, CTrans \cite{7}, CDur \cite{19}, the envelope-based SED approach \cite{6}, and the Multi-Branch model \cite{4}. As shown in Table~\ref{tab:sed_results}, our model achieves a segment-level F-score of 83.15\% and a class-wise accuracy of 95.17\%, yielding a nearly 7\% relative improvement over the strongest baseline, TSDNet (76.3\%). This gain highlights the benefit of the unified encoder, which learns a single consistent representation for reference matching and event detection without relying on separate branches or handcrafted duration priors.

\begin{figure}[ht]
    \centering
    \includegraphics[width=\linewidth,height=5.0cm]{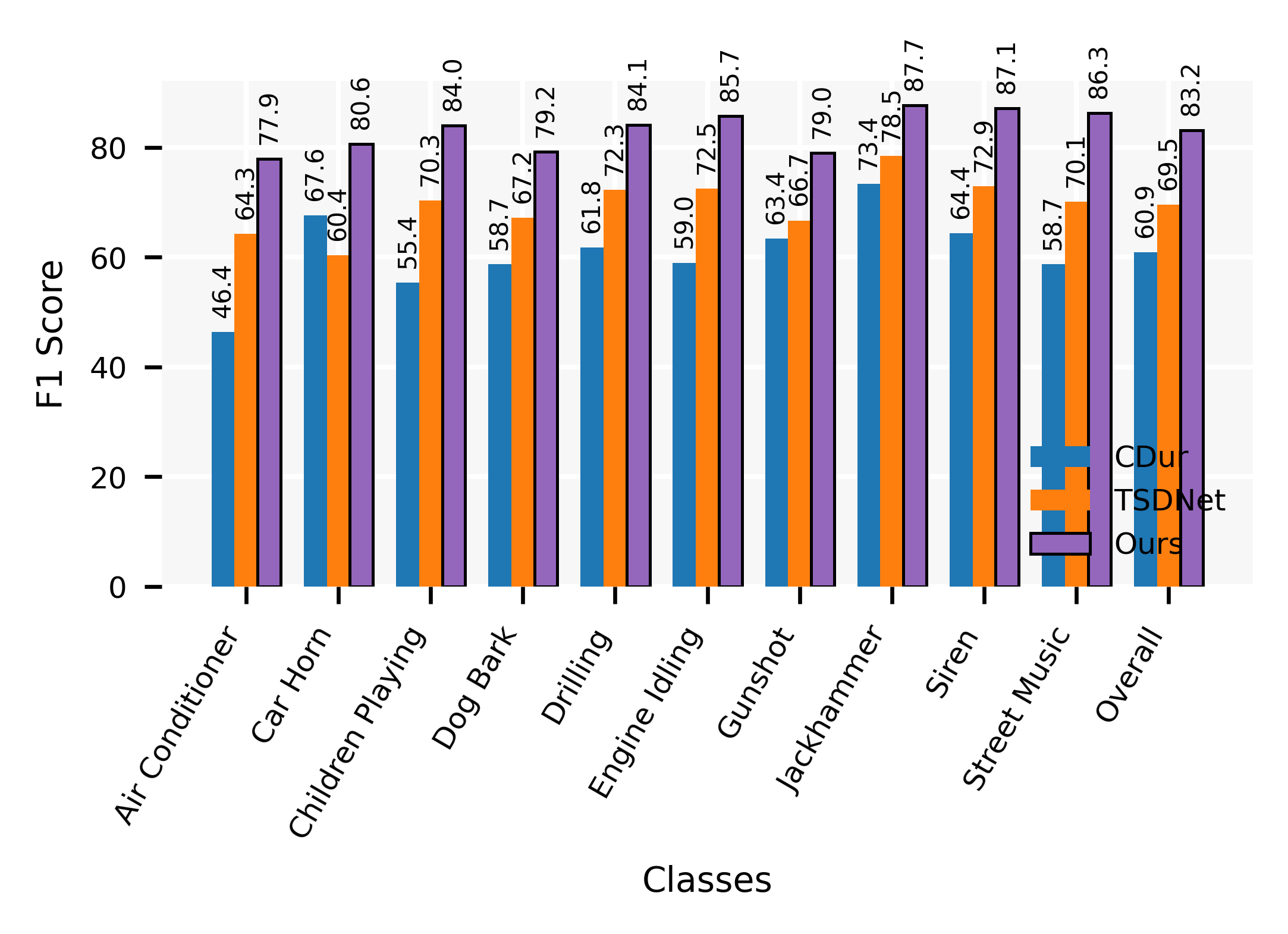}
    \captionsetup{font=small}
    \caption{Comparison of per-class F1 scores on URBAN-TSD-Strong, reproduced under the same evaluation protocol.}
        
    \label{fig:UrbanStrong_perclass}
\end{figure}

Figure~\ref{fig:UrbanStrong_perclass} reports per-class F1 scores on URBAN-SED for CDur, TSDNet, and our model. Across nearly all ten classes, our approach yields higher F1 scores, with particularly pronounced improvements for transient events such as \emph{Car Horn}, \emph{Dog Bark}, and \emph{Gunshot}. Although some variation remains between classes, reflecting inherent differences in event duration, spectral overlap, and background complexity, the overall pattern confirms that richer feature representations and the proposed architecture provide more robust detection across diverse urban sounds. To evaluate generalization beyond the training label set, we train the model on 7 URBAN-SED classes and evaluate it on the full 10-class test set. The model achieves 73.47\% segment-level F1 and 91.06\% accuracy, indicating only modest degradation despite excluding three classes during training. On the unseen classes (Jackhammer, Siren, and Street Music), the model attains segment-level accuracies of 86.0\%, 85.2\%, and 86.0\%, compared to 96.0\%, 96.1\%, and 95.8\% when trained on all classes. \\

\begin{table}[]
\centering
\renewcommand{\arraystretch}{0.9} 
\captionsetup{font=small}
\caption{Segment-based F1 (\%) and Accuracy (\%) of baseline TSD models and the proposed method. Accuracy is reported only for models re-trained under identical settings (*); other results are from original papers.}
\small
\begin{tabular}{l|c|c}
\hline
\textbf{Method} & \textbf{  Segment-based F1} & \textbf{Accuracy} \\
\hline
Multi-Branch \cite{4}    & 61.60 & --     \\
CDur \cite{19}*             & 64.75 & 90.03     \\
Supervised SED \cite{6}    & 64.70 & --     \\
CTrans \cite{7}                 & 65.14 & --     \\
TSDNet \cite{3}*                                     & 76.3 & 90.77  \\
\hline
\textbf{Unified (Ours)}                     & \textbf{83.15} & \textbf{95.17} \\
\hline
\end{tabular}

\label{tab:sed_results}
\end{table}

\begin{table}[]
\captionsetup{font=small}
\caption{Performance comparison of unified and dual-branch encoder designs on URBAN-SED (dual-branch follows the TSDNet \cite{3} detection network).}
\small
\begin{tabular}{c|c|c|c} \hline
\textbf{Backbone}                  & \textbf{Encoder Design} & \textbf{Segment-F1} & \textbf{Accuracy} \\ \hline
\multirow{2}{*}{\textbf{CNN14}}    & Dual-branch             & 71.19               & 91.27             \\
                                   & Unified                 & 74.20               & 91.66             \\ \hline
\multirow{2}{*}{\textbf{ConvNeXt}} & Dual-branch             & 80.38               & 93.81             \\
                                   & Unified                 & 83.15               & 95.17  \\            \hline
\end{tabular}

\label{tab:unified vs dual}
\end{table}

\noindent\textbf{Encoder Design Comparison:} Across ConvNeXt and CNN14 backbones, the unified encoder consistently outperforms the dual-branch design (Table \ref{tab:unified vs dual}), highlighting the benefit of shared representation learning for reference-guided TSD. \\

\noindent\textbf{Effect of Fusion Strategy:} We compared different reference–mixture fusion strategies within the unified encoder. Element-wise multiplication achieves segment-level F1 of 83.15\%, providing strong and efficient baseline. FiLM-based \cite{perez2017filmvisualreasoninggeneral} conditioning improves performance to 83.18\%, while cross-attention yields the best result with 86.06\% segment-level F1, by enabling content-adaptive alignment between reference and mixture representations, allowing the model to selectively emphasize target-relevant mixture features. \\

\noindent\textbf{Cross-Domain Evaluation on AudioSet-Strong:} As described in Section \ref{label:dataset}, we evaluated our model trained on URBAN-TSD-Strong on a curated AudioSet-Strong subset to test cross-domain generalization. Despite the domain shift, the model achieves an F1 score of 76.62\%, showing strong performance across diverse classes (Table \ref{tab:tsd_strong_perclass}). These results highlight the robustness of our unified encoder in handling real-world acoustic variability. While the ConvNeXt encoder’s pretraining on AudioSet may aid transfer, the results reflect the strength of our architecture and training strategy. \\

\begin{figure}[h]
    \centering
    \includegraphics[width=\linewidth,height=3.5cm]{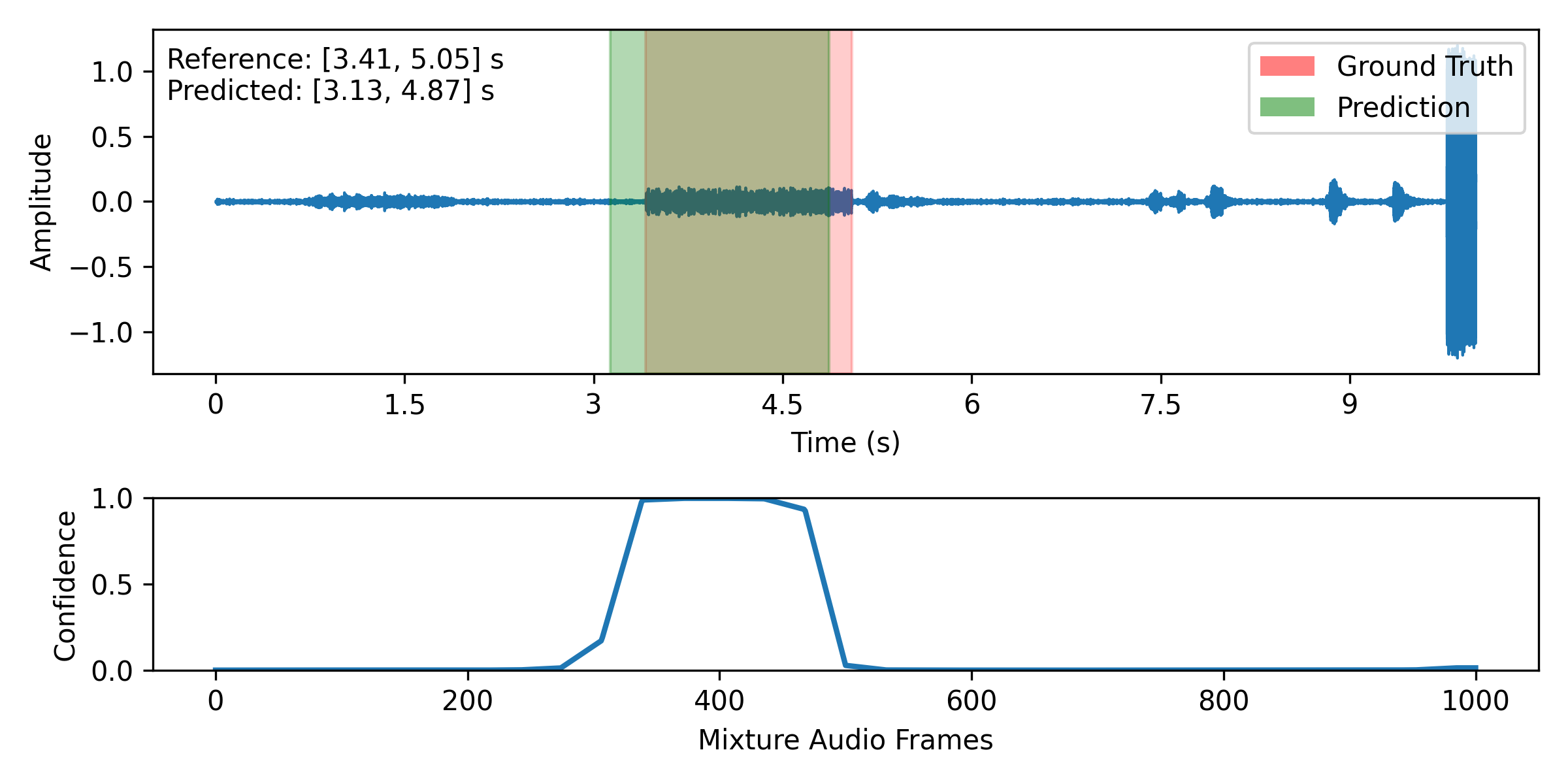}
     \captionsetup{font=small}
    
    \caption{Temporal localization example depicting waveform with ground-truth and predicted event boundaries, together with the model’s frame-level confidence scores.}
    \label{fig:localization}
\end{figure}

\noindent\textbf{Temporal Localization and Confidence Analysis:} To qualitatively assess temporal localization, we visualize the waveform with ground-truth and predicted boundaries and overlay the model’s frame-level confidence scores. As shown in Figure~\ref{fig:localization}, predicted boundaries closely align with the ground truth, and confidence peaks correspond well with the target event region. This demonstrates the model’s ability to localize events precisely, even in real-world mixtures. \\

\noindent\textbf{Impact of Urban-TSD-Strong+:} Training on Urban-TSD-Strong+ reduces segment-level F1 from 83.15\% to 78.94\% compared to Urban-TSD-Strong, as the model must additionally reject absent-class references, making the task harder but more realistic . Incorporating contrastive setup may help mitigate this effect and is left for future exploration.

\begin{table}[!t]
\centering
\renewcommand{\arraystretch}{0.7} 
\captionsetup{font=small}
\caption{Cross-domain evaluation on AudioSet-Strong showing per-class segment-based F1 scores across diverse acoustic events.}
\small
\begin{tabular}{|l|c|c|}
\hline
\multirow{2}{*}{\textbf{Classes}} & \multicolumn{2}{c|}{\textbf{AudioSet Strong}} \\
\cline{2-3}
& \textbf{F1 (\%)} & \textbf{Accuracy (\%)} \\
\hline
Air conditioner  & 74.6 & 99.4 \\
Car horn         & 64.1 & 97.4 \\
Children playing & 74.4 & 97.2 \\
Dog bark         & 77.5 & 98.3 \\
Drilling         & 86.1 & 98.6 \\
Engine idling    & 71.8 & 90.4 \\
Gun shot         & 70.6 & 97.3 \\
Jackhammer       & 59.5 & 97.9 \\
Siren            & 92.0 & 98.1 \\
Street music     & 95.7 & 98.8 \\
\hline
\textbf{Average} & \textbf{76.6} & \textbf{97.3} \\
\hline
\end{tabular}

\label{tab:tsd_strong_perclass}
\end{table}

\section{Conclusion}
\label{ssec:Conclusion}
We present a unified encoder framework for reference-guided targeted sound detection, leveraging a single shared network to jointly process mixture and reference audio. This design simplifies the model while improving feature alignment and generalization. Trained with strong supervision in a multi-task setup, our approach achieves state-of-the-art performance on URBAN-SED and generalizes effectively to real-world recordings in AudioSet-Strong. Ablation studies further confirm its robustness in temporal localization and resilience to distributional shifts, establishing a strong benchmark for open-domain acoustic retrieval.

  


 


\clearpage

\bibliographystyle{IEEEbib}
\bibliography{strings,refs}

@inproceedings{1,
  author    = {Justin Salamon and Christopher Jacoby and Juan Pablo Bello},
  title     = {A Dataset and Taxonomy for Urban Sound Research},
  booktitle = {Proceedings of the 22nd ACM International Conference on Multimedia (ACM MM)},
  year      = {2014},
  pages     = {1041--1044},
  doi       = {10.1145/2647868.2655045}
}

@inproceedings{2,
  author    = {Justin Salamon and Duncan MacConnell and Mark Cartwright and Peter Li and Juan Pablo Bello},
  title     = {Scaper: A library for soundscape synthesis and augmentation},
  booktitle = {2017 IEEE Workshop on Applications of Signal Processing to Audio and Acoustics (WASPAA)},
  year      = {2017},
  pages     = {344--348},
  doi       = {10.1109/WASPAA.2017.8170052}
}

@inproceedings{3,
  author    = {Helin Wang and Dongchao Yang and Yuexian Zou and Fan Cui and Yujun Wang},
  title     = {Detect What You Want: Target Sound Detection},
  booktitle = {Proceedings of the Detection and Classification of Acoustic Scenes and Events (DCASE) Workshop},
  year      = {2022},
  month     = nov,
  address   = {Nancy, France},
  pages     = {3--4},
}

@inproceedings{4,
  author    = {Y. Huang and X. Wang and L. Lin and H. Liu and Y. Qian},
  title     = {Multi-Branch Learning for Weakly-Labeled Sound Event Detection},
  booktitle = {ICASSP 2020 - 2020 IEEE International Conference on Acoustics, Speech and Signal Processing (ICASSP)},
  year      = {2020},
  pages     = {641--645},
  publisher = {IEEE},
  doi       = {10.1109/ICASSP40776.2020.9054033},
  url       = {http://arxiv.org/abs/2002.09661}
}

@inproceedings{6,
  author    = {I. Martin-Morato and A. Mesaros and T. Heittola and T. Virtanen and M. Cobos and F. J. Ferri},
  title     = {Sound Event Envelope Estimation in Polyphonic Mixtures},
  booktitle = {ICASSP 2019 - IEEE International Conference on Acoustics, Speech and Signal Processing (ICASSP)},
  year      = {2019},
  pages     = {935--939},
  publisher = {IEEE},
  doi       = {10.1109/ICASSP.2019.8682377}
}

@inproceedings{7,
  author    = {K. Miyazaki and T. Komatsu and T. Hayashi and S. Watanabe and T. Toda and K. Takeda},
  title     = {Weakly-supervised sound event detection with self-attention},
  booktitle = {ICASSP 2020 - 2020 IEEE International Conference on Acoustics, Speech and Signal Processing (ICASSP)},
  year      = {2020},
  pages     = {66--70},
  publisher = {IEEE},
  doi       = {10.1109/ICASSP40776.2020.9054565}
}

@inproceedings{8,
  author    = {Jort F. Gemmeke and Daniel P. W. Ellis and Dylan Freedman and Aren Jansen and Wade Lawrence and R. Channing Moore and Manoj Plakal and Marvin Ritter},
  title     = {AudioSet: An Ontology and Human-Labeled Dataset for Audio Events},
  booktitle = {Proceedings of the IEEE International Conference on Acoustics, Speech and Signal Processing (ICASSP)},
  year      = {2017},
  pages     = {776--780},
  doi       = {10.1109/ICASSP.2017.7952261}
}

@misc{9,
  author       = {Thomas Pellegrini and Ismail Khalfaoui-Hassani and Etienne Labb{\'e} and Timoth{\'e}e Masquelier},
  title        = {Adapting a ConvNeXt model to audio classification on AudioSet},
  year         = {2023},
  note         = {Accepted at INTERSPEECH 2023; arXiv preprint arXiv:2306.00830},
  eprint       = {2306.00830},
  archivePrefix= {arXiv},
  primaryClass = {cs.SD}
}

@inproceedings{10,
  author    = {Dongchao Yang and Helin Wang and Zhongjie Ye and Yuexian Zou and Wenwu Wang},
  title     = {RaDur: A Reference-Aware and Duration-Robust Network for Target Sound Detection},
  booktitle = {Proceedings of Interspeech 2022},
  year      = {2022},
  pages     = {433--437},
  doi       = {10.21437/Interspeech.2022-433},
  url       = {https://arxiv.org/abs/2204.02143}
}

@misc{11,
      title={Target Speaker Extraction through Comparing Noisy Positive and Negative Audio Enrollments}, 
      author={Shitong Xu and Yiyuan Yang and Niki Trigoni and Andrew Markham},
      year={2025},
      eprint={2502.16611},
      archivePrefix={arXiv},
      primaryClass={cs.SD},
      url={https://arxiv.org/abs/2502.16611}, 
}

@misc{12,
      title={FlowTSE: Target Speaker Extraction with Flow Matching}, 
      author={Aviv Navon and Aviv Shamsian and Yael Segal-Feldman and Neta Glazer and Gil Hetz and Joseph Keshet},
      year={2025},
      eprint={2505.14465},
      archivePrefix={arXiv},
      primaryClass={eess.AS},
      url={https://arxiv.org/abs/2505.14465}, 
}

@misc{14,
      title={Improving Target Sound Extraction with Timestamp Information}, 
      author={Helin Wang and Dongchao Yang and Chao Weng and Jianwei Yu and Yuexian Zou},
      year={2022},
      eprint={2204.00821},
      archivePrefix={arXiv},
      primaryClass={cs.SD},
      url={https://arxiv.org/abs/2204.00821}, 
}

@article{15,
  author    = {Phat Lam and Lam Pham and Dat Tran and Alexander Schindler and Silvia Poletti and Marcel Hasenbalg and David Fischinger and Martin Boyer},
  title     = {Aud-Sur: An audio analyzer assistant for audio surveillance applications},
  journal   = {arXiv preprint arXiv:2503.23827},
  year      = {2025},
  url       = {https://arxiv.org/abs/2503.23827}
}

@article{16,
  author    = {Andreea-Maria Oncescu and A. Sophia Koepke and João F. Henriques and Zeynep Akata and Samuel Albanie},
  title     = {Audio Retrieval with Natural Language Queries},
  journal   = {arXiv preprint arXiv:2105.02192},
  year      = {2021},
  url       = {https://arxiv.org/abs/2105.02192}
}

@article{17,
  author    = {Khairunisa Sharif and Bastian Tenbergen},
  title     = {Smart Home Personal Assistants: A Security and Privacy Review},
  journal   = {ACM Computing Surveys (CSUR)},
  volume    = {52},
  number    = {3},
  pages     = {1--36},
  year      = {2019},
  doi       = {10.1145/3293663},
  url       = {https://dl.acm.org/doi/10.1145/3293663}
}

@inproceedings{18,
  author    = {Dongchao Yang and Helin Wang and Wenwu Wang and Yuexian Zou},
  title     = {A Mixed Supervised Learning Framework for Target Sound Detection},
  booktitle = {Proceedings of Interspeech 2020},
  year      = {2020},
  pages     = {861--865},
  doi       = {10.21437/Interspeech.2020-861},
  url       = {https://www.isca-speech.org/archive/Interspeech_2020/yang20i_interspeech.html}
}

@article{19,
author = {Dinkel, Heinrich and Wu, Mengyue and Yu, Kai},
title = {Towards Duration Robust Weakly Supervised Sound Event Detection},
year = {2021},
issue_date = {2021},
publisher = {IEEE Press},
volume = {29},
issn = {2329-9290},
url = {https://doi.org/10.1109/TASLP.2021.3054313},
doi = {10.1109/TASLP.2021.3054313},
journal = {IEEE/ACM Trans. Audio, Speech and Lang. Proc.},
month = jan,
pages = {887–900},
numpages = {14}
}

@article{20,
author = {Kong, Qiuqiang and Xu, Yong and Wang, Wenwu and Plumbley, Mark D.},
title = {Sound Event Detection of Weakly Labelled Data With CNN-Transformer and Automatic Threshold Optimization},
year = {2020},
issue_date = {2020},
publisher = {IEEE Press},
volume = {28},
issn = {2329-9290},
url = {https://doi.org/10.1109/TASLP.2020.3014737},
doi = {10.1109/TASLP.2020.3014737},
journal = {IEEE/ACM Trans. Audio, Speech and Lang. Proc.},
month = aug,
pages = {2450–2460},
numpages = {11}
}

@article{21,
author = {Kong, Qiuqiang and Yu, Changsong and Xu, Yong and Iqbal, Turab and Wang, Wenwu and Plumbley, Mark D.},
title = {Weakly Labelled AudioSet Tagging With Attention Neural Networks},
year = {2019},
issue_date = {November 2019},
publisher = {IEEE Press},
volume = {27},
number = {11},
issn = {2329-9290},
url = {https://doi.org/10.1109/TASLP.2019.2930913},
doi = {10.1109/TASLP.2019.2930913},
journal = {IEEE/ACM Trans. Audio, Speech and Lang. Proc.},
month = nov,
pages = {1791–1802},
numpages = {12}
}

@Article{22,
AUTHOR = {Mesaros, Annamaria and Heittola, Toni and Virtanen, Tuomas},
TITLE = {Metrics for Polyphonic Sound Event Detection},
JOURNAL = {Applied Sciences},
VOLUME = {6},
YEAR = {2016},
NUMBER = {6},
ARTICLE-NUMBER = {162},
URL = {https://www.mdpi.com/2076-3417/6/6/162},
ISSN = {2076-3417},
DOI = {10.3390/app6060162}
}

@misc{perez2017filmvisualreasoninggeneral,
      title={FiLM: Visual Reasoning with a General Conditioning Layer}, 
      author={Ethan Perez and Florian Strub and Harm de Vries and Vincent Dumoulin and Aaron Courville},
      year={2017},
      eprint={1709.07871},
      archivePrefix={arXiv},
      primaryClass={cs.CV},
      url={https://arxiv.org/abs/1709.07871}, 
}

\end{document}